\documentclass[twocolumn,twocolappendix]{aastex7}
\usepackage{amsmath}
\usepackage{amssymb}
\begin{document}

\title{The Role of Magnetospheric Rebound in Breaking Resonant Chains of Super-Earths and Mini-Neptunes}

\author[orcid=0000-0002-0162-163X]{Mengrui Pan}
\affiliation{Department of Earth, Environmental and Planetary Sciences, MS 126, Rice University, Houston, TX 77005, USA}
\affiliation{Rice Space Institute, MS 126, Rice University, Houston, TX 77005, USA}
\email{mp183@rice.edu}  

\author[orcid=0000-0003-1878-0634]{Andre Izidoro}
\affiliation{Department of Earth, Environmental and Planetary Sciences, MS 126, Rice University, Houston, TX 77005, USA}
\email{ai18@rice.edu}

\author[orcid=0000-0002-5418-6336]{Sho Shibata}
\affiliation{Department of Earth, Environmental and Planetary Sciences, MS 126, Rice University, Houston, TX 77005, USA}
\email{ss508@rice.edu}

\begin{abstract}
Stellar magnetic fields are thought to truncate the inner regions of protoplanetary disks around T Tauri stars, creating a magnetospheric cavity near the star. As the disk evolves and disperses, the truncation radius is expected to move outward as the balance between magnetic and viscous forces shifts. Planets migrating inward can become trapped near the inner edge, but as the edge itself moves outward, the evolving disk torques can drive planets to migrate outward as well. We employ N-body simulations to assess the influence of magnetospheric cavity expansion on the dynamical evolution and orbital architectures of compact resonant chains of super-Earths and mini-Neptunes. Our results show that rebound-driven expansion of the disk's inner edge plays a pivotal role in destabilizing resonant chains by spreading planetary systems outward, thereby triggering early dynamical instabilities and giant impacts. Despite this dynamical evolution, key observable properties of close-in planetary systems --such as the distribution of orbital period ratio, the intra-system similarity in planet sizes (``radius uniformity''), and the bimodal distribution of planet radii known as the ``radius valley''-- remain largely consistent with those of systems formed without the rebound effect, in which the inner edge of the disk remains fixed. Thus, the primary consequence of the rebound appears to be the early disruption of resonant chains, rather than any significant alteration to the statistical properties of the resulting super-Earth and mini-Neptune populations.
\end{abstract}

\keywords{\uat{Exoplanet systems}{484} --- \uat{Dynamical evolution}{421}  --- \uat{Orbital resonances}{1181} --- \uat{Super Earths}{1655} --- \uat{Mini Neptunes}{1063}}

\section{Introduction} 

Super-Earths and mini-Neptunes, planets with physical radii, or size,  ranging from 1 to $4 \ R_{\oplus}$ and orbital periods shorter than $100$ days, are common around Sun-like stars \citep{Fressin2013, Petigura2013, Marcy2014}. Observations have placed strong constraints on several of their key properties, including the period ratio distribution \citep{Lissauer2011, Fabrycky2014}, the uniformity of planet sizes within individual systems \citep{Millholland2017, Weiss2018}, and their bimodal size distribution \citep{Fulton2017, Fulton&Petigura2018}. Notably, the distribution of period ratios between adjacent planets shows a broad spread, indicating that most planet pairs are not trapped in first-order mean motion resonances \citep{Lissauer2011, Fabrycky2014}.

Driven by disk-planet interactions, super-Earths and mini-Neptunes are thought to undergo orbital migration within the gaseous disk \citep{Goldreich&Tremaine1979} during their formation phase. The early-formed planets typically stop migration near the inner edge of the disk \citep{ Romanova2019}, followed by the inward migration of subsequent planets \citep{Chatterjee&Tan2014, Ormel2017}. During the migration phase, planets are typically captured into mean motion resonances, naturally leading to the formation of compact resonant configurations and long resonant chains \citep{Gillon2017, Izidoro2017, Charalambous2018, Izidoro2021}. The overproduction of resonance contrasts sharply with the observed population, highlighting the necessity of further post-formation resonance-breaking mechanisms \citep{Lissauer2011, Fabrycky2014}.

Several scenarios have been proposed so far to explain the discrepancies between observations and theoretical expectations, including inside-out dispersal of the gaseous disk \citep{Liu2017, Liu&Ormel2017, Liu2022}, planet-planet interactions \citep{Izidoro2017, Izidoro2021, Izidoro2022}, interaction with planetesimals \citep{Chatterjee&Ford2015}, long-term mass loss \citep{Matsumoto&Ogihara2020, Wang&Lin2023}, tidal dissipation \citep{Delisle2012, Delisle2014}, and overstable liberation \citep{Goldreich&Schlichting2014, Hands&Alexander2018, Li2024}. 
Yet, the extent to which some of these mechanisms influence the overall architecture of planetary systems and the statistical distribution of super-Earths/mini-Neptunes remains unclear.

In this paper, we concentrate on the effect of the inside-out disk dispersal, driven by magnetospheric rebound of the disk inner edge, on the late-stage dynamical evolution of planetary systems. The role of this process has been investigated in the context of idealized 2-planet systems considering varying parameters such as planet mass, disk accretion rate, stellar magnetic field strength, and disk dispersal timescale \citep{Liu2017, Liu&Ormel2017}. 
Other studies have focused instead on testing the importance of this process for specific systems, such as Trappist-1 \citep{Pichierri2024}.
Here we adopt a more comprehensive approach by analyzing a suite of multi-planet systems produced in planet formation simulations that incorporate key physical processes \citep{Shibata&Izidoro2025}, such as planetesimal and pebble accretion, gas-driven migration, and disk evolution. We compare outcomes from simulations that include the rebound mechanism with those that neglect it, by keeping the inner edge fixed.

This paper is organized as follows. Section \ref{sec:model} describes the physical model and initial conditions. Section \ref{sec:results} presents the main simulation results, followed by a comparison of different scenarios in Section \ref{sec:comp}. In Section \ref{fig:obs}, we examine the observational characteristics of the resulting planetary systems. Finally, Section \ref{sec:summary} summarizes our findings and discusses their implications.

\section{Model} \label{sec:model}

\subsection{Planet Formation  Model} \label{sec:setup}

In this work, we follow the formation model developed by \citet{Shibata&Izidoro2025}. \citet{Shibata&Izidoro2025} run N-body simulations of two rings of planetesimals: one located between 0.5-1.5 au and another between 8-15 au. In their model, growth in the inner ring is dominated by planetesimal accretion, while planets in the other disk grow primarily via pebble accretion. The planets formed from the inner planetesimal ring become rocky super-Earths, and those from the outer planetesimal ring become icy mini-Neptunes. They set the total mass of the inner planetesimal ring and the gaseous disk's lifetime as free parameters, and run 50 simulations for each parameter set by using slightly different distributions of planetesimals. They found that the architectures and orbital/physical properties of super-Earths and mini-Neptunes can be broadly reproduced when the inner disk's mass is $6 M_\oplus$ and the disk's lifetime is $2$ Myrs (labeled \textit{M6T2} in their model).

%\citet{Shibata&Izidoro2025} showed that the architectures and orbital/physical properties of super-Earths and mini-Neptunes can be broadly reproduced when planets form from two rings of planetesimals: one located between 0.5-1.5 au and another between 8-15 au. In their model, growth in the inner ring is dominated by planetesimal accretion, while planets in the other disk grow primarily via pebble accretion. The total mass in the inner disk is treated as a free parameter, and their results suggest that typical systems form in disks with $\lesssim 6 \ M_{\oplus}$ in solids. 

%In this work, we adopt the simulations of \citet{Shibata&Izidoro2025} to model the dynamical evolution of resonant chains undergoing magnetospheric rebound. We use their most successful set of simulations that better match observations---namely their model where the inner disk mass is $6 \ M_{\oplus}$ and the disk dispersal timescale is 2 Myr (their model \textit{M6T2}). 

To model the dynamical evolution of resonant chains undergoing magnetospheric rebound, we adopt their \textit{M6T2} simulations. 
Instead of simulating planet formation from the beginning of the disk phase, we extract planetary system configurations from their simulations at the stellar age of 2 Myr. This epoch corresponds to the onset of the disk dispersal, by which point most systems have established compact resonant chains near the disk inner edge, at this point located at 0.1 au. When extracting planetary systems from these simulations, we include only planets more massive than 0.1 $M_{\oplus}$. This threshold does not affect the quality of our results but allows us to significantly reduce computational time, as many simulations contain numerous low-mass leftover objects in the other disk.

Using these pre-formed planetary systems, we perform numerical simulations using the hybrid symplectic/Bulirsch-Stoer integrator in the \textit{Mercury} N-body code \citep{Chambers1999}, focusing on the impact of magnetospheric rebound on the orbital evolution and statistical properties of super-Earth and mini-Neptune populations. We consider disk dispersal timescales of $t_{\rm disk} = 5 \times 10^4$, $10^5$, and $5 \times 10^5\ \rm yr$, along with a control model excluding the rebound effect (where the disk inner edge is assumed to be static). For each scenario, we run 50 simulations and integrate each system up to 50 Myr.

The simulations of \citet{Shibata&Izidoro2025} provide both the masses and bulk compositions of planets, which are categorized as rocky or water-rich, with or without atmospheric envelopes. For mass-radius conversion, we use the composition-dependent relationships of \citet{Zeng2019}. We assume an initial atmospheric mass fraction of 0.003 \citep{Owen&Wu2017, Izidoro2022}. During the gas-rich phase, we adopt perfect merging. After disk dispersal, we follow \citet{Biersteker&Schlichting2019} and assume that giant impacts with a projectile-to-target mass ratio $M_{\rm pro}/M_{\rm tar} > 0.1$ lead to complete atmosphere loss. We also incorporate atmospheric erosion due to photoevaporation driven by stellar irradiation, following the prescription of \citep{Misener&Schlichting2021}.

\subsection{Disk model} \label{sec:dm}

Our simulations, which account for the effect of magnetospheric rebound at the inner edge of the disk, begin with resonant chains extracted from \citet{Shibata&Izidoro2025}. These systems are reinserted into a gaseous disk that reflects the final disk conditions at 1.9 Myr in their models. The disk structure is identical to that adopted in \citet{Shibata&Izidoro2025}, following a standard one-dimensional steady-state $\alpha-$disk prescription. The surface density profile is given by the classical formulation of \citet{Shakura&Sunyaev1973}
\begin{equation}
    \Sigma_{\rm g} = \frac{\dot{M}_{\rm g}}{3 \pi \nu},
\end{equation}
where $\dot{M}_{\rm g}$ and $\nu=\alpha H_{\rm g}^2 \Omega_{\rm K}$. We set the dimensionless viscosity $\alpha=0.004$. $\Omega_{\rm K}$ is the Keplerian angular velocity, and $H_{\rm g}$ is the disk scale height. 

The radial profile of the disk is based on 3D hydrodynamical simulations \citet{Bitsch2015}, and the disk accretion rate and stellar age correlate as
\begin{equation}
    \log \left(\frac{\dot{M}_{\rm g}}{M_{\odot} \, \rm yr^{-1}} \right) = -8 - 1.4 \log \left( \frac{t_{\rm d, 0} + 10^5 \, \rm yr}{10^6 \, \rm yr} \right).
\end{equation}
We initialize the disk at $t_{\rm d, 0} = 1.9 \ \rm Myr$, when the disk accretion rate $\dot{M}_{\rm g, 0} \simeq 3.79 \times 10^{-9} \ M_{\odot} \ \rm{yr^{-1}}$. By this time, the resonant configurations of the planetary systems have largely formed, and the disk subsequently begins to dissipate exponentially 
\begin{equation}
    \dot{M}_{\rm g} = \dot{M}_{\rm g, 0} \exp \left( -\frac{t-t_0}{t_{\rm disk}} \right),
\end{equation}
where $t$ represents the stellar age. $t_0 = t_{\rm d, 0} + 10^5  \ \rm yr$, and $t_{\rm disk}$ is the disk depletion timescale---a free parameter that we vary across values of $5 \times 10^4$, $10^5$ (fiducial), and $5 \times 10^5$ yr. 
%These different depletion timescales are consistent with varying stellar magnetic field strengths. Adopting this range of depletion timescales allows us to account---at least indirectly---for the effects of stellar magnetic fields, which we do not model explicitly. 
Importantly, the values of $t_{\rm disk}$ assumed here are roughly a factor of 10-100 shorter than typical disk dispersal timescales \citep{Mamajek2009, Williams&Cieza2011}. This choice is intentional, as our simulations begin at 1.9 Myr, by which point the disk is already substantially depleted.

The disk temperature and aspect ratio are assumed to remain constant over time and can be well fitted by power laws \citep{Bitsch2015}
\begin{equation}
    T_{\rm g} = 164.6 \left( \frac{r}{1 \ \rm AU} \right)^{-0.91} \ \rm K, 
\end{equation}
\begin{equation}
    h_{\rm g} = 0.026 \left( \frac{r}{1 \ \rm AU} \right)^{0.05},
\end{equation}
where $r$ is the distance to the central star.  

We model the evolution of the disk inner edge assuming that the protoplanetary disk is truncated by the stellar magnetic field, at the location where the Lorentz torque balances the viscous torque \citep{Ghosh&Lamb1979, Koenigl1991}. 
% As the disk disperses, the viscous torque diminishes. Gas near the cavity is accreted onto the central star along magnetic field lines, resulting in the outward shift of the disk inner boundary. 
The location of the disk inner edge is given by \citep{Frank2002, Armitage2010, Liu2017} 
\begin{equation}
\begin{aligned}
   &r_{\rm c} = \left( \frac{B_{\star}^4 R_{\star}^{12}}{4 G M_{\star} \dot{M}_{\rm g}^2} \right)^{1/7} \\
    &= 0.1 \left( \frac{\dot{M}_{\rm g}} {3.79 \times 10^{-9} \ M_{\odot} \ \rm yr^{-1}} \right)^{-2/7} \left( \frac{B_{\star}}{1.95 \ \rm kG} \right)^{4/7} \ \rm au, 
\end{aligned} 
\end{equation}
where $B_{\star}=1.95 \ \rm{kG} $, $R_{\star}$, and $M_{\star}$ are stellar magnetic field strength, radius, and mass, respectively. At the beginning of our simulations, the disk inner edge is located at $r_{\rm c, \, 0} = 0.1$ au. As the disk disperses over time, the viscous torque---which scales with $\dot{M}_{\rm g}$---diminishes, resulting in the outward shift of the disk inner boundary. We recall that $t_{\rm disk}$ governs how rapidly the disk disperses in our model and consequently how fast the disk's inner edge moves outward during the dispersal phase.

\subsection{Type I torques} \label{sec:torque}

Planets embedded in the disk and sufficiently distant from the disk inner edge have their migration direction and speed determined by the total disk torque, which includes the two-sided (inner and outer) Lindblad torque ($\Gamma_{\rm L, 2s}$) and corotation torque ($\Gamma_{\rm c, 2s}$) \citep{Tanaka2002, Paardekooper2010, Paardekooper2011}. Assuming a local isothermal disk, the total torque can be written as \citep{Paardekooper2010}
\begin{equation}
    \begin{split}
        &\frac{\Gamma_{\rm 2s}}{m_{\rm p} (r_{\rm p} \Omega_{\rm p})^2} = \frac{\Gamma_{\rm L, \, 2s} + \Gamma_{\rm c, \, 2s}}{m_{\rm p} (r_{\rm p} \Omega_{\rm p})^2} = \\
        &\left[ -(2.5 + 0.5 \beta + 0.1 s) + 1.4 \beta + 1.1 \left( \frac{3}{2} +s \right) \right] q_{\rm d} \frac{q_{\rm p}}{h_{\rm g}^2},
    \end{split}
\end{equation}
where $\beta = \partial \ln T_{\rm g} / \partial \ln r$ and $s = \partial \ln \Sigma_{\rm g} / \partial \ln r$ is the gradient of temperature and gas surface density, respectively. $q_{\rm d} \equiv \Sigma_{\rm p} r_{\rm p}^2 / M_{\star}$, and $q_{\rm p} = M_{\rm p}/ M_{\star}$ is the planet-star mass ratio.

For planets near the inner edge of the protoplanetary disk, only one-sided torques operate to drive their migration due to the absence of inner disk \citep{Liu2017, Liu&Ormel2017}. In particular, the Lindblad torque from the outer disk applies a net negative torque on the planet, driving inward migration. In the corotation region, gas is transported exclusively from the exterior orbit to the interior orbit. The upper branch of the horseshoe flow exerts a positive torque on the planet, promoting outward migration. The magnitude of one-sided torque can be expressed as \citep{Lin&Papaloizou1993, Paardekooper&Papaloizou2009, Liu2017}
\begin{equation}
    \frac{\Gamma_{\rm 1s}}{m_{\rm p} (r_{\rm p} \Omega_{\rm p})^2} = \Gamma_{\rm L, \, 1s} + \Gamma_{\rm c, \, 1s} = C_{\rm L} q_{\rm d} \frac{q_{\rm p}}{h_{\rm g}^3} + C_{\rm hs} q_{\rm d} \frac{q_{\rm p}}{h_{\rm g}^3},
\end{equation}
where $\Gamma_{\rm L, \, 1s}$ and $\Gamma_{\rm c, \, 1s}$ represent the one-sided Lindblad torque and one-sided corotation torque, respectively. $C_{\rm L} = -0.65$, and $C_{\rm hs} = 2.46$.

We use $f = \exp[-(r-r_{\rm c})/x_{\rm hs}]$ as a smooth function to connect the one-sided and two-sided torque regions, where $x_{\rm hs} = 1.7 (q_{\rm p}/h_{\rm g})^{0.5} r_{\rm p}$ is the half-width of the horse-shoe region \citep{Paardekooper&Papaloizou2009, Ormel2013}. The combined migration torque is given by
\begin{equation}
    \Gamma = 
    \begin{cases}
        f\Gamma_{\rm 1s} + (1-f)\Gamma_{\rm 2s}, & \text{when } r \geq r_{\rm c} \\
        0, & \text{when } r < r_{\rm c}.
    \end{cases}
\end{equation}

We then obtained the planetary migration rate 
\begin{equation}
    \frac{\dot{a}}{a} = \frac{2 \Gamma}{m_{\rm p} r_{\rm p}^2 \Omega_{\rm p}}.
\end{equation}

Besides, planetary eccentricity and inclination are damped by disk tidal effect, with a characteristic damping timescale of \citet{Cresswell&Nelson2008}
\begin{equation}
    \begin{aligned}
        t_{\rm e} = &\frac{t_{\rm wave}}{0.780} \times \Bigl[ 1-0.14 \left( \frac{e}{h_{\rm g}} \right)^2 \\
        & + 0.06 \left( \frac{e}{h_{\rm g}} \right)^3 + 0.18 \left( \frac{e}{h_{\rm g}} \right) \left( \frac{i}{h_{\rm g}} \right)^2 \Bigl]
    \end{aligned}
\end{equation}
\begin{equation}
    \begin{aligned}
        t_{\rm i} = &\frac{t_{\rm wave}}{0.544} \times \Bigl[ 1-0.30 \left( \frac{i}{h_{\rm g}} \right)^2 \\
        & + 0.24 \left( \frac{i}{h_{\rm g}} \right)^3 + 0.14 \left( \frac{e}{h_{\rm g}} \right)^2 \left( \frac{i}{h_{\rm g}} \right) \Bigl]
    \end{aligned}
\end{equation}
where $t_{\rm twave} = h_{\rm g}^4 q_{\rm p}^{-1} q_{\rm d}^{-1}  \Omega^{-1}$.

Planets located inside the disk's inner cavity are assumed to reside in a gas-depleted region, and therefore do not experience significant migration torques.

% We then simulated disk dissipation with different timescales $t_{\rm disk} = 5 \times 10^4$, $10^5$ and $5 \times 10^5 \ \rm yr$ to explore how disk sweeping speed influences the population distribution of super-Earths and mini-Neptunes.
% We conducted 50 simulations for each disk model and integrated them up to 50 Myr to systematically analyze the difference among the planet populations.

\section{Results} \label{sec:results}

% The scenario with disk dissipation timescale of $t_{\rm disk} = 10^5\ \rm yr$, termed the fiducial case, is described in Sec.~\ref{sec:fid}. We compared the results of different disk lifetime in Sec.~\ref{sec:tau}. In Sec.~\ref{sec:noreb}, we investigate the scenario where disk inner boundary is fixed. 
% We then compare the orbital distribution and physical properties of planets in Sec.~\ref{sec:comp}.

% \subsection{Fiducial case} \label{sec:fid}

\begin{figure*}[ht!]
    \includegraphics[width=\textwidth]{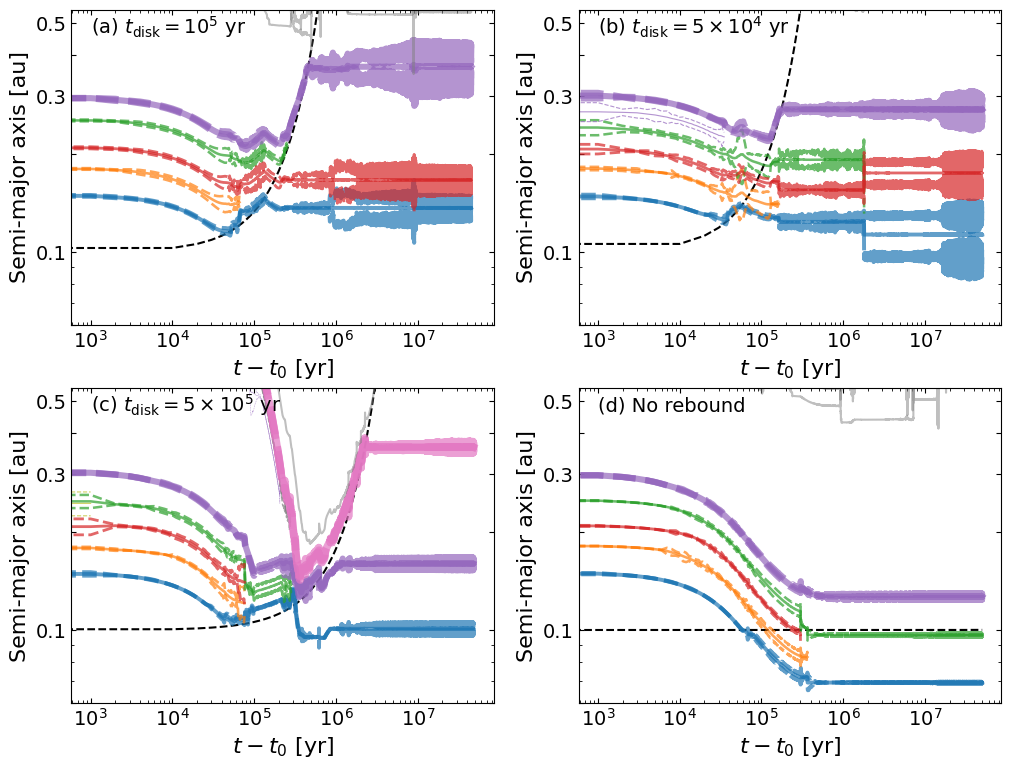}
    \caption{The semi-major axis evolution of planets with orbital period shorter than $100$ days in the disk with different dissipation timescale of $t_{\rm disk} = 10^5$ (fiducial case, panel a), $t_{\rm disk} = 5 \times 10^4$ (b), $t_{\rm disk} = 5 \times 10^5$ (c), and $t_{\rm disk} = 10^5$ (panel d). In panel (a-c), the rebound effect is included, whereas in panel (d), the disk undergoes only exponentially dissipation without cavity expansion over time. Solid colored lines show the evolution of planets’ semi-major axes. Perihelion and aphelion are shown in lighter shades of the same color. The black dashed lines in each panel indicate the disk inner edge. A stronger rebound effect leads to a more widely separated planetary system.
    }
    \label{fig:a_comp}
\end{figure*}

In Fig. \ref{fig:a_comp}a, we show the evolution of planets in one of our simulations with a disk depletion timescale of $t_{\rm disk} = 10^5$ yr (our fiducial scenario). This figure displays the temporal evolution of the planet's semi-major axes as the disk begins to dissipate exponentially.
% the planetary system during the dispersal of the gas disk, where the perihelion and aphelion are shown in dashed lines. 
The initial magnetospheric cavity\footnote{Throughout the paper, we use ``cavity'' and ``inner edge'' interchangeably to refer to the innermost extent of the gas disk set by the stellar magnetosphere.} is located at 0.1 au and expands outward during the dissipation phase, as indicated by the black dashed line. In this case, twenty planets more massive than $0.1 \ M_{\oplus}$ are initially presented between approximately 0.1 au and 10 au, but only those with final orbital periods shorter than 100 days are shown for presentation purposes. 

% Since the rebound effect diminishes with disk dispersal, the gaseous disk is nearly depleted when the cavity expands to the orbital period of $P = 100 \ \rm d$. Planets in the outer regions of the protoplanetary disk rarely undergo the rebound effect and primarily serve as gravitational perturbers on the inner super-Earths and mini-Neptunes. 

Before describing the evolution of planets during the rebound of the inner edge, we recall that the planetary systems are extracted from the simulations of \citet{Shibata&Izidoro2025} at 1.9 Myr. At this time, in some systems, planets may not yet have reached the disk's inner edge, while in others they may already be located there. In addition, at the beginning of our simulations, planets can still migrate inward because the gas disk remains relatively dense. Even in systems where the innermost planets are temporarily trapped at the inner edge, they can be pushed into the cavity by external planets that are part of---or that subsequently join---the resonant chain.

Fig.~\ref{fig:a_comp}a shows that planets migrate slightly inward at the beginning of the simulation, driven by two-sided torques. As the system approaches the cavity edge, the innermost planet becomes subject to one-sided torques. The positive horseshoe drag overwhelms the negative Lindblad torque, driving the planet to migrate outward. Meanwhile, outer planets, far away from the inner disk edge, are subject to two-sided torques and migrate inward. This convergent migration compacts the planetary configuration and quickly triggers dynamical instabilities at approximately $t-t_0 = 70 \ \rm Kyr$.

% Since the sum of the torques on all planets is dominated by the one-sided torque from the inner planet, planets migrate outward. 
% The planet reverses its migration direction and moves outward.
% it continuously exchanges angular momentum with the outer planets through gravitational interactions, driving the entire resonant chain to migrate outward altogether. During this stage, the resonance configuration may become increasingly compact, potentially triggering dynamical instability. 

As the gaseous disk continues to dissipate, the timescale for (outward) planetary migration eventually becomes longer than the timescale for the recession of the disk's inner edge. When this happens, the near-edge planet in Fig.~\ref{fig:a_comp}a (blue line) is left behind inside the expanding cavity at 150 kyr. This evolution allows the now-second innermost planet (red line in Fig.\ref{fig:a_comp}a ) to resume inward migration which pushes the innermost planet (blue line) further in due to their resonant interaction. Eventually, as the inner edge continues to expand, the red-line planet undergoes a similar evolution to that of the innermost planet, being left inside the disk cavity. The overall system architecture becomes more radially spread out during the sweeping of the inner edge with dynamical instabilities and collisions taking place.

By the time the expanding inner cavity reaches approximately 0.42 au, corresponding to an orbital period of $P = 100$ d, the gas surface density is already very depleted. Planets located in the outer regions of the protoplanetary disk feel a relatively much weaker rebound effect (compared to those close-in), yet they act as gravitational perturbers on the interior planets. In the subsequent gas-free phase, long-term dynamical interactions---particularly with the distant and typically icy planets (see \citet{Shibata&Izidoro2025})---gently excite the eccentricities and inclinations of the inner super-Earths and mini-Neptunes. As shown in Fig.~\ref{fig:a_comp}a, the innermost two planets remain on near-circular orbits when they are left in the disk cavity. At $t-t_0 \simeq 5\times10^5 \ \rm yr$ and $10^7$ yr, respectively, the eccentricities of these planets are moderately excited, reaching $\sim 0.1 - 0.2$, under the perturbation of outer disk planets (solid lines in gray).

In Fig.~\ref{fig:a_comp}b and c, we show the dynamical evolution of planets in two simulations with disk depletion timescales of $t_{\rm disk} = 5 \times 10^4$ yr and $t_{\rm disk} = 5 \times 10^5$ yr, respectively.

In the rapidly dissipating disk (Fig.~\ref{fig:a_comp}b), planets experience only slight outward migration before being quickly stranded inside the cavity. Their orbits remain more tightly packed compared to the fiducial case. The original resonance chain is disrupted by a violently dynamical instability inside the cavity at approximately 1.8 Myr, which strongly excites the eccentricities of the surviving planets (see pericenter and apocenter lines). Under the joint gravitational disturbances from the outer planets, the remnant three planets end with eccentricities of 0.12, 0.17, and 0.25 (inner to outer), slightly larger than those in Fig.~\ref{fig:a_comp}a.

In contrast to Fig.~\ref{fig:a_comp}b, slower dissipation (Fig.~\ref{fig:a_comp}c) permits more sustained inward migration and promotes the formation of more compact system architectures. Unlike the case with shorter dissipation timescales, an icy planet with a mass of $5.69 \ M_{\oplus}$---initially located at approximately $1.9 \ \rm au$---migrates deep into the inner disk and triggers a dynamical instability (pink line in panel c). Because collisions mainly occur while the disk is still gas-rich, eccentricities and inclinations excited during the instability are efficiently damped. The strong rebound associated with the long-lived gas disk drives significant orbital expansion and helps preserve the system’s near-coplanarity and near-circularity ($i \lesssim 2^\circ$, $e \lesssim 0.05$) throughout its evolution. As we show in the next section, the orbital inclination distribution in this long-dissipation scenario differs markedly from that of systems with shorter dissipation timescales (e.g., the fiducial case).

Fig.~\ref{fig:a_comp}d shows the evolution of planets in a disk with $t_{\rm disk} = 10^5 \ \rm yr$, neglecting the effects of magnetospheric rebound. In this classical planetary migration scenario---where the inner edge of the disk remains fixed (e.g. \citet{Shibata&Izidoro2025})---planets migrate inward under the influence of two-sided torques, and the innermost ones are often pushed into the cavity by resonant interactions with exterior companions. The outermost three planets are captured into a 5:4-5:4 resonant chain. As the gaseous disk dispersal, the first-order resonant chain is disrupted due to gravitational instability. Eventually, the surviving planets settle into near-circular ($e \lesssim 0.02$) and near-coplanar ($i \lesssim 5^\circ$) orbits that could remain dynamically stable for at least $\sim 50 \ \rm Myr$. Note that this represents just one example from our suite of simulations. This type of evolution is consistent with the range of outcomes seen in previous simulations of the breaking-the-chain model \citep{Izidoro2021, Shibata&Izidoro2025}, in which first-order resonant chains formed through migration in a gaseous disk are generally destabilized by gravitational interaction.

As shown by the simulations in Fig.~\ref{fig:a_comp}a-c, enhanced rebound contributes to the divergence of planetary orbits. Consequently, in the early stage of dynamical evolution, the stronger the rebound effect, the lower the planetary eccentricities and inclinations. However, in some cases, planets in the outer disk act as gravitational perturbers, destabilizing the orbits of the inner planets over long-term evolution and exciting their eccentricities and inclinations. Thus, the direct effect of rebound on planetary orbital properties is difficult to isolate, but a statistical comparison is provided in Sec.~\ref{sec:ei}.

% In this case, planetary migration is governed exclusively by two-sided torques. The innermost planet migrates inward and halts near the inner cavity. It is subsequently captured into a resonance chain with outer planets and pushed further inward. Dynamical instability is triggered at $t-t_0 \simeq 3 \times 10^5 \ \rm yr$ once the disk has been largely depleted. Eventually, the system settles with three close-in super-Earths, locked in 3:2 and 5:3 mean motion resonances, respectively.

% Driven by the continued slow expansion of the disk cavity, planetary orbits expand significantly during outward migration. 
% , which give rise to mutual collisions (at $t-t_0 \simeq 80 \ \rm Kyr$ in Fig.~\ref{fig:a_comp}c)
% The latecomer collides with the planet indicated by the green line, scatters the innermost planet inward, and is subsequently cleared by the disk and relocated to an outer orbit.

\section{Comparing planetary architectures} \label{sec:comp}

In this section, we compare the orbital architectures of planetary systems formed under different disk dissipation scenarios, analyzing the outcomes of 50 simulations for each case.

Our results show that the rebound of the disk's inner edge can significantly increase the likelihood that a planetary system becomes dynamically unstable during or after the gas disk phase. We define a system as dynamically unstable if, by the end of the simulation, it retains at most one pair of planets in mean-motion resonance. 

% planetary populations generated by 50 simulations for each disk model is analyzed in this section. Numerical results indicate that a longer disk depletion timescale promotes wider orbital separation during cavity expansion, which enhanced the likelihood of system destabilization and resonance disruption (Fig. \ref{fig:a_comp}c). 

For rapidly depleted disks ($t_{\rm disk} = 5 \times 10^4 \ \rm yr$), resonance chains are disrupted in 88\% of systems. This fraction increases to 92\% in the fiducial case ($t_{\rm disk} = 10^5 \ \rm yr$), and to 94\% in long-lived disk scenario ($t_{\rm disk} = 5 \times 10^5 \ \rm yr$). In contrast, only 64\% of systems experience resonance disruption in the absence of magnetospheric rebound. Note that Each series of simulations utilizes the same initial conditions. Further details on the final resonant structure of systems formed under different scenarios are provided in Appendix \ref{app:sys}.

To better isolate the role of rebound, we further classify the non-rebound outcomes into two categories based on their dynamical evolution after disk dispersal: (i) stable systems, which remain ``resonant'' (containing at least two planet pairs in first-order mean motion resonance), and (ii) unstable systems, which undergo dynamical instabilities (less than two planet pairs in first order mean motion resonance). These two subsets are compared with the full set of systems formed under rebound scenarios\footnote{Since the majority of rebound cases result in dynamical instability, the inclusion of a few stable systems does not significantly affect the overall statistics.}. Interestingly, the fraction of unstable systems in the rebound cases is consistent with that found to best reproduce observed planetary system architectures in the ``breaking the chains'' model \citep{Izidoro2017, Izidoro2021}.

Throughout our analysis, we restrict the sample to planets with orbital periods shorter than 100 days, in line with the observed population of inner super-Earths and mini-Neptunes.
% In following analysis, we categorize the outcomes of non-rebound cases into two groups based on their post-disk dynamical evolution: stable systems without dynamical instability after disk dissipation and unstable systems with it. Both groups are compared with full set of systems formed in scenarios incorporating the rebound effect \footnote{Given that most cases in these scenarios are dynamically unstable, the present of stable systems does not affect the overall results.}. We also note that our analysis exclusively includes planets with orbital periods shorter than 100 days.

% We compare the period ratio (Sec. \ref{sec:pr})
% with preserved resonance chains

\subsection{Period ratio} \label{sec:pr}

\begin{figure*}
    \includegraphics[width=\textwidth]{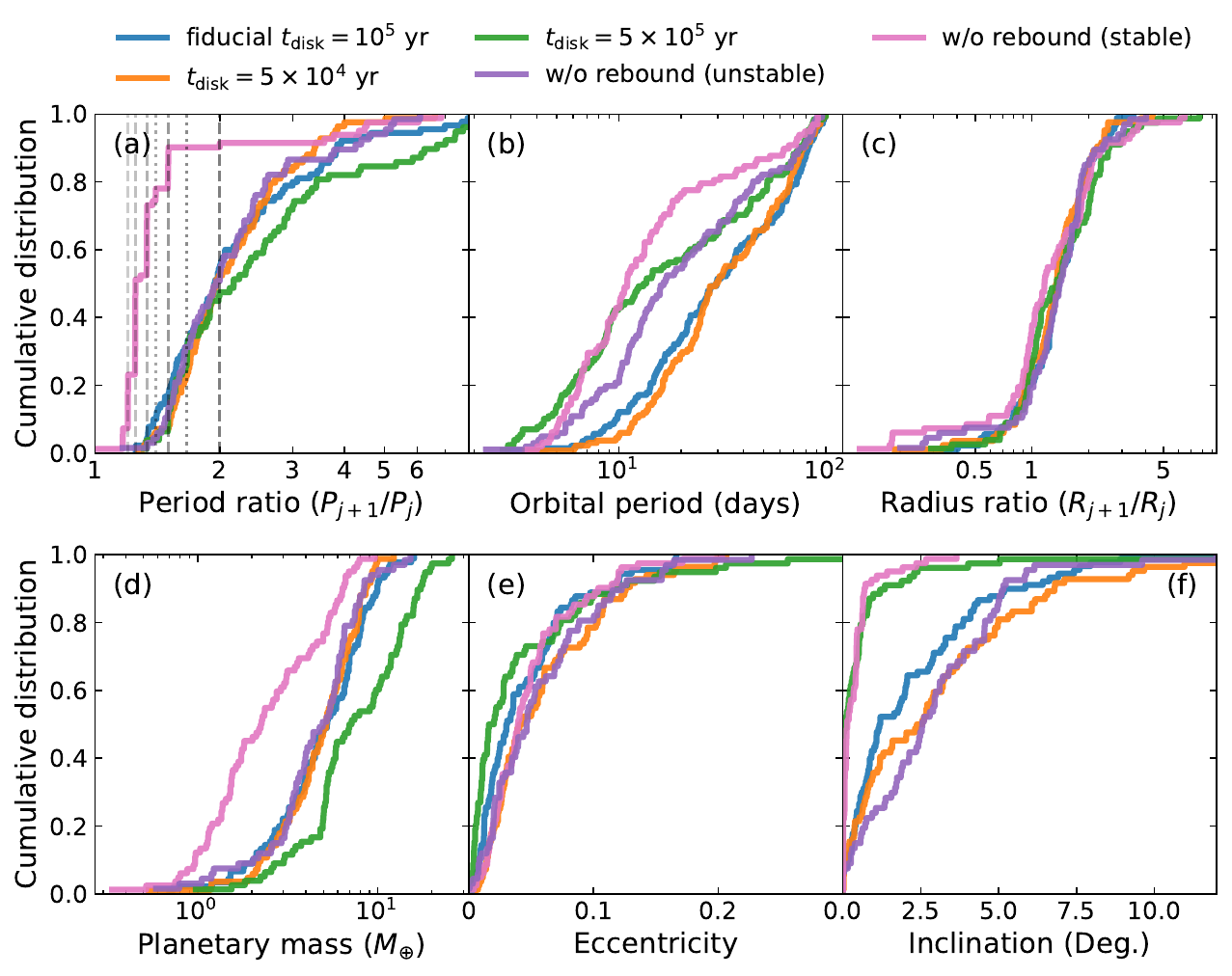}
    \caption{
    Cumulative distributions of period ratio (panel a), orbital period (panel b), radius ratio (panel c), planetary mass (panel d), eccentricity (panel e), and inclination (panel f). Colored lines correspond to planet populations formed in different disk models. Dashed lines in panel (a) indicate first-order resonances, while dash-dotted lines represent second-order resonances. The outcomes of the no-rebound scenario is categorized into stable and unstable system groups. The rebound effect efficiently disrupts the original resonance chains established in the gaseous disk without substantially altering the statistical distributions of planet populations.
    \label{fig:models}}
\end{figure*}

The cumulative period ratio distribution of adjacent planets pairs is shown in Fig. \ref{fig:models}a. We first compare stable and unstable systems from the non-rebound scenario to isolate the role of dynamical instability. Planets in stable systems (pink line) frequently occupy first-order mean-motion resonances---such as 6:5, 5:4, 4:3, 3:2, and 2:1 (black dashed lines)---as well as second-order resonances like 5:3 and 7:5 (black dotted lines). 

By contrast, dynamically unstable systems from the non-rebound scenario (purple line) exhibit a broader distribution of period ratios and lack prominent resonance features. Interestingly, this distribution closely resembles those produced in simulations that include the magnetospheric rebound effect (orange, blue, and green lines). Among these, systems formed in lone-lived disks (green line) show slightly wider separations between adjacent planets, consistent with a stronger and more sustained rebound effect. Despite these subtle differences, pairwise Kolmogorov-Smirnov (KS) tests between the various models yield $P$-values greater than 0.17, indicating that the magnetospheric rebound does not significantly affect the overall period ratio distribution of close-in super-Earths and mini-Neptunes when compared to the classical scenario in which the disk's inner edge remains static.

\subsection{Period distribution} \label{sec:pd}

In Fig. \ref{fig:models}b, the orbital period distributions vary significantly across models. The distribution for the fiducial cases closely resembles that of the short-lived disk model, with a KS test $p$-value of approximately 0.45. However, in systems formed in long-lived disks (slower rebound), planets tend to accumulate either very close to the star ($P<10 \ \rm d$) or at wider separations ($P > 30 \ \rm d$). This bimodal distribution arises from the more sustained inward migration of outer icy planets at early times, which can push or scatter the innermost planets into the disk cavity. As the inner edge of the disk later rebounds outward, the remaining planets are driven to wider orbits.

In non-rebound scenarios, planets are also progressively shepherded inward by resonant interactions with outer migrating companions. As a result, nearly 80\% of planets in stable systems end up with orbital periods shorter than 20 days. In contrast, unstable systems show broader and more dispersed period distributions, primarily due to collisions and mergers that occurred after gas dispersal. Notably, the more extended inward migration phase permitted in simulations with longer disk dispersal timescales plays a key role in shaping these trends.

\subsection{Planet radius and mass} \label{sec:rm}

Kepler observations have revealed that planets within multi-planet systems tend to have similar sizes to their immediate neighbors---a phenomenon known as the ``peas in a pod'' pattern \citep{Weiss2018, Weiss2023}. Fig. \ref{fig:models}c shows the cumulative distribution of radius ratios between adjacent planets, defined as immediate orbital neighbors in the system with no other planets between them in semi-major axis, in our simulations. All scenarios exhibit a strong clustering around unity, indicating a high degree of size uniformity among neighboring planets, regardless of the disk dissipation history. 

In contrast, the planetary mass distribution shown in Fig.~\ref{fig:models}d reveals noticeable variations in the total mass of planetary systems, particularly among stable systems and those formed in long-lived disks. Stable systems tend to preserve their initial low-mass configurations, resulting in a shallower slope in the cumulative mass distribution. In unstable systems, however, frequent collisions and mergers driven by gravitational interactions significantly increase planetary masses. The fiducial scenario, the rapidly depleted disk case, and unstable non-rebound systems exhibit statistically indistinguishable mass distributions. By comparison, systems formed in long-lived disks show systematically higher total masses, resulting from a larger number of collisions and the rapid inward migration of outer mini-Neptunes.

\subsection{Radius valley distribution} \label{sec: rv}

Observations indicate that super-Earths and mini-Neptunes follow a bimodal size distribution, with two peaks near 1.4 and 2.4 $R_{\oplus}$, and a valley around 1.8 $R_{\oplus}$---a feature commonly referred to as the ``radius valley'' \citep{Fulton2017, Fulton&Petigura2018}. A widely supported interpretation of this valley is that it reflects a dichotomy in planetary bulk composition: planets on the left side of the valley (super-Earths) are predominantly rocky, while those on the right side (mini-Neptunes) are rich in volatiles, especially water-ice \citep{Zeng2019, Otegi2020, Venturini2020, Izidoro2022}. 

\begin{figure*}
    \includegraphics[width=\textwidth]{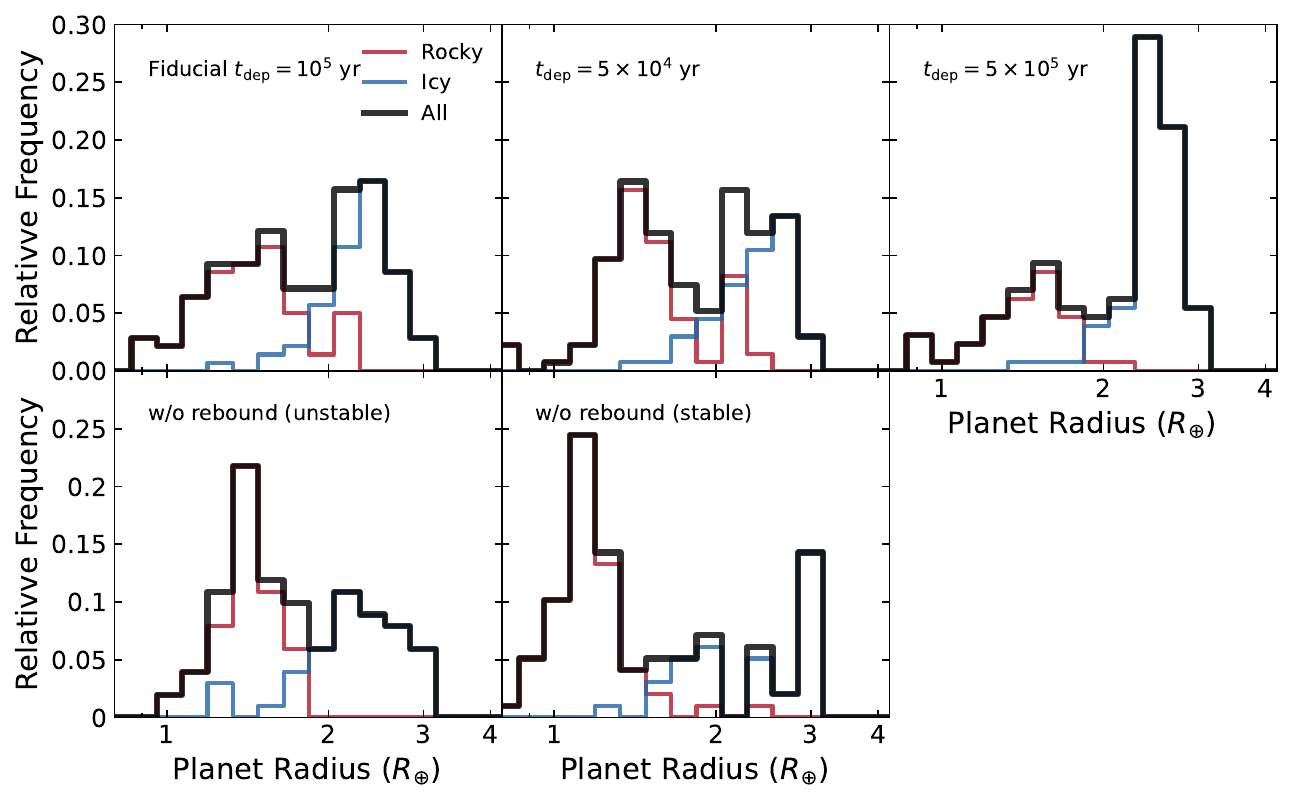}
    \caption{Radius distributions of super-Earths and mini-Neptunes in different disk models. The results for dynamically unstable systems in each model, along with stable systems formed in disks without the rebound effect, are illustrated. The rebound effect efficiently disrupts the original resonance chains established in the gaseous disk without substantially altering the statistical distributions of planet populations.
    \label{fig:RV}}
\end{figure*}

We compared the radius distribution of planets from different scenarios in Fig. \ref{fig:RV}. The red and blue curves correspond to the rocky and icy planets, respectively, while the black curve shows the combined population. 
In simulations that include magnetospheric rebound or result in dynamical instability, a clear peak-valley-peak structure emerges, with a pronounced valley between $\sim 1.84-2.05 \ R_{\oplus}$ and peaks at $\sim 1.48{-}1.65 \ R_{\oplus}$ and $2.28-2.55 \ R_{\oplus}$. The locations of these features remain relatively consistent across models, with rocky planets primarily populating the left side of the valley, and icy planets the right. 

The Key difference across scenarios lies in the relative abundance of rocky versus icy planets. As disk lifetimes increase, a larger number of icy planets migrate inward and populate the inner regions, increasing their overall representation. Meanwhile, these inward-migrating icy planets can trigger additional collisions among inner rocky planets, reducing the number of rocky survivors. Therefore, in simulations with short-lived disks ($ t_{\rm disk} \leq 10^5 \ \rm yr$), rocky and icy planets are present in roughly equal numbers. In contrast, in long-lived disk scenarios ($ t_{\rm disk} = 5 \times 10^5 \ \rm yr$), icy planets dominate the inner planetary population. 

Interestingly, systems that remain dynamically stable throughout the simulation do not exhibit a clear peak-valley-peak structure. In these systems, the overall mass and compositional diversity are lower, leading to a smoother and less structured radius distribution.

\subsection{Eccentricity and Inclination} \label{sec:ei}

The cumulative distributions of planet eccentricities and inclinations across different scenarios are shown in Fig. \ref{fig:models}e and f. As expected, dynamically unstable systems exhibit higher eccentricities and inclinations compared to stable ones due to gravitational perturbations and collisions during late-stage instabilities. This trend is consistent with the results of previous simulations \citep{Izidoro2021}. In the case of long-lived disks, the stronger and more sustained rebound effect drives planets to wider orbital separations, which in turn reduces the likelihood of post-disk instabilities. As a result, systems formed in long-lived disks tend to retain relatively low inclinations, closely resembling the distribution observed in stable systems. Overall, planets are more likely to maintain low-eccentricity, coplanar orbits when the rebound effect is strong and sustained. 

% triggers more intense instabilities at early stage and 
% However, most instabilities occur within gaseous disks that eccentricities and inclinations are rapidly damped by tidal interaction. In addition,

\subsection{Multiplicity} \label{sec:mul}

\begin{figure}
    \includegraphics[width=\columnwidth]{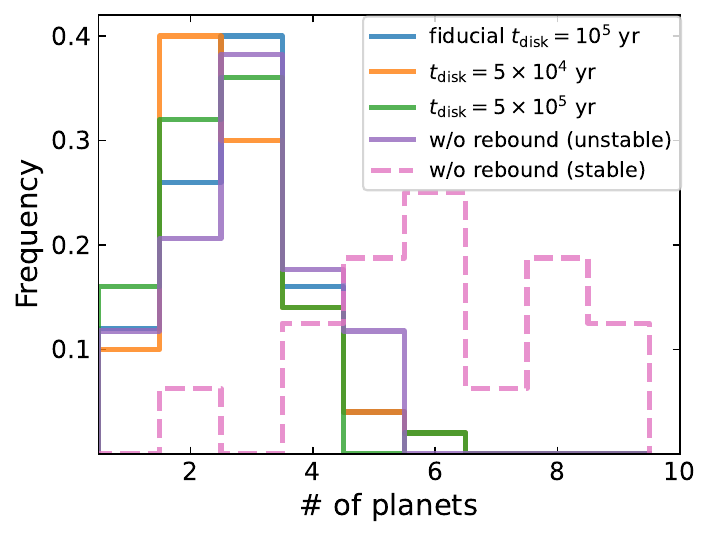}
    \caption{Number of planets with orbital period shorter than 100 days in planetary systems formed in different models. Planet multiplicity distributions are consistent across rebound scenarios and unstable non-rebound systems, with each system typically holding 2–3 planets. By contrast, stable non-rebound systems are characterized by higher multiplicity, usually comprising 4 to 8 planets.
    \label{fig:multi}}
\end{figure}

Fig. \ref{fig:multi} shows the intrinsic planet multiplicity of systems formed under different scenarios. With the exception of stable systems, all other scenarios yield broadly similar multiplicity distributions. Approximately 10\% of systems host a single super-Earth or mini-Neptune with an orbital period $P \leq 100 \ \rm d$, about 65\% systems harbor two or three planets, and roughly 20\% contain more than three. To quantitatively assess these similarities, we performed two-sample KS tests between all pairs of scenarios. The resulting $p$-values are generally high (above 0.48), indicating no statistically significant differences in planet multiplicity. 
% The one exception is the comparison between the short-lived disk model and unstable non-rebound systems, which yields the lowest $p$-value of approximately 0.48. Nonetheless, this still does not provide strong evidence to reject the null hypothesis that the distributions are drawn from the same underlying population.

% that the distributions of planet multiplicity in different scenarios are statistically consistent.

In contrast, stable systems exhibit systematically higher multiplicities. Most of these systems host more than three planets, with an average number of approximately six. Planets in these systems are commonly locked in resonant chains, maintaining dynamically stable orbital configurations. 

The similarity between unstable non-rebound systems and those formed in rebound scenarios suggests that magnetospheric rebound acts primarily to destabilize planetary system architectures, rather than altering their overall statistical multiplicity distributions.

\subsection{Resonant chain breaking epoch}

\begin{figure}
    \includegraphics[width=\columnwidth]{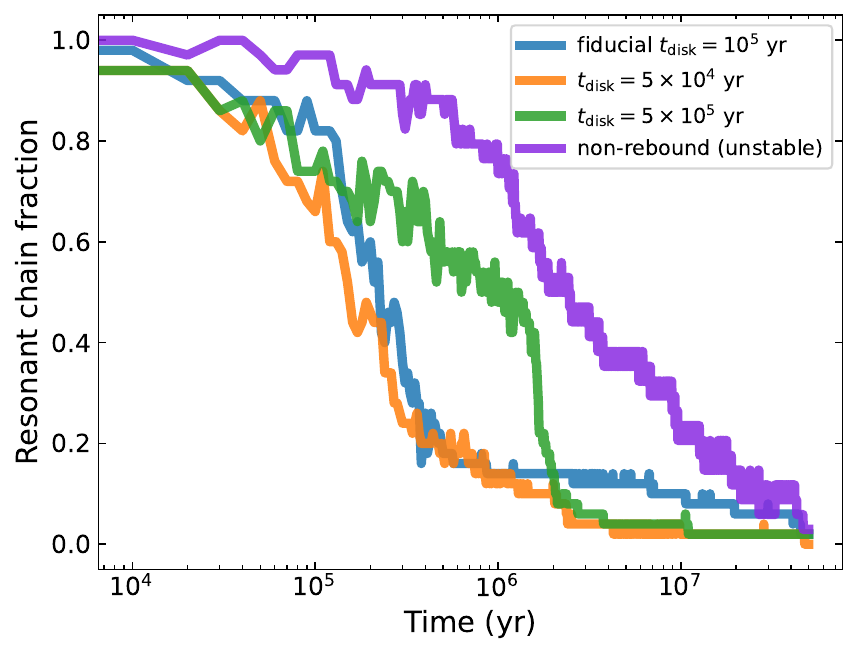}
    \caption{Temporal evolution of resonant chain disruption for $P < 100 \ \mathrm{d}$ under various scenarios. 
    \label{fig:FC}}
\end{figure}

Fig. \ref{fig:FC} shows the fraction of systems maintaining resonant chains as a function of time, in the context of orbital period with $P < 100$ d, across different models. 
We recall that the resonant chain configuration refers to a system where at least three planets are locked in resonance. Resonant chains in systems formed under rebound scenarios are generally broken in the very early stage when the disk is still gas-rich. In the long-lived disk scenario, a pronounced peak appears at around $3 \times 10^5$ yr, coinciding with the timescale over which icy planets from the outer disk migrate inward, reaching the inner planets. This peak corresponds to the inward migration of icy planets from the outer disk, which can interact with and rebuild resonant chains with inner planets. This is not observed in the other cases because icy planets do not make it into the inner disk ($P < 100$ days) due to the relatively faster disk dispersal.

In unstable systems formed without magnetospheric rebound, resonant chain disruption occurs over relatively longer timescales compared to rebound cases, with most events beginning only after disk dispersal (around 0.4~Myr). For example, in the scenario with a short disk lifetime ($t_{\rm disk} = 5 \times 10^4$~yr), approximately 50\% of resonant chains are disrupted within the first 0.1~Myr of the simulation. In contrast, the same level of disruption is reached only after $\sim$2~Myr in the non-rebound case. We note that the final fraction of stable systems in the non-rebound model is significantly higher than in scenarios that include the rebound. Our results suggest that the magnetospheric rebound effect is the primary driver of very early dynamical instabilities, leading to the disruption of resonant chains during the gas-rich phase of disk evolution.

% However, no evident peak-valley-peak structure is discovered in the stable systems.

\section{Comparing simulations and observations} \label{sec:obs}

To compare the observable properties of our simulated planetary systems with real exoplanet observations, we perform synthetic transit observations following the framework of \citet{Izidoro2017}. Viewing geometries are uniformly sampled in azimuthal from $0^\circ$ to $360^\circ$ and vertically from $30^\circ$ above to $30^\circ$ below the midplane ($i=0^\circ$), using a grid of 360 azimuthal and 600 vertical points. 

\begin{figure*}[ht!]
\includegraphics[width=\textwidth]{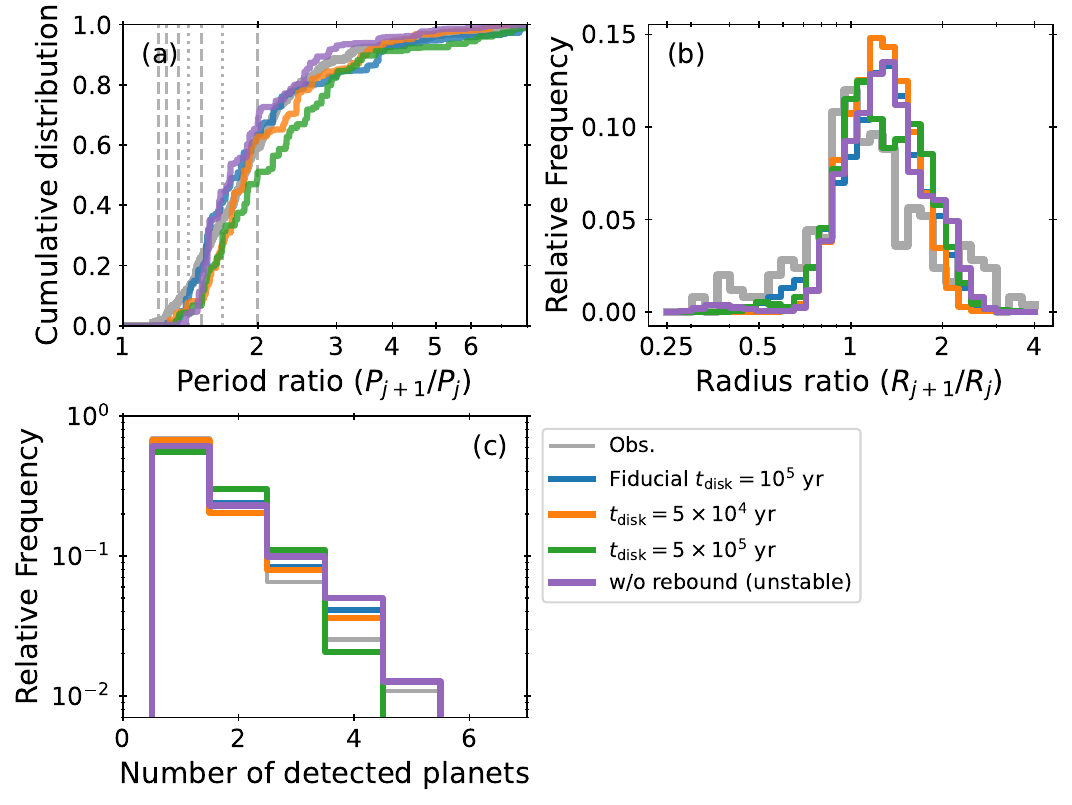}
\caption{Comparison of observationally simulated period ratio (panel a), radius ratio (panel b), and multiplicity (panel c) distributions from different models (colored lines) and the California Kepler Survey (CKS) dataset (gray lines). Same as those in Fig. \ref{fig:models}a, the vertical gray dashed and dotted lines indicates the first- and second-order resonances, respectively.
\label{fig:obs}}
\end{figure*}

\begin{figure*}[htbp!]
\includegraphics[width=\textwidth]{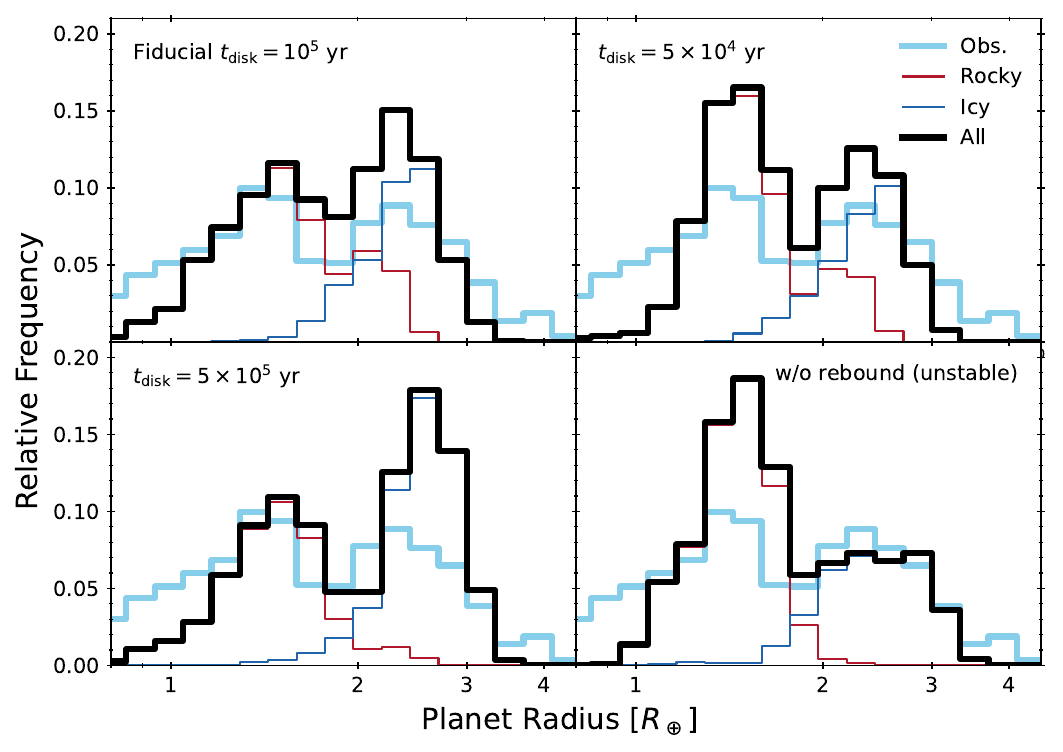}
\caption{Same as Fig. \ref{fig:RV}, but showing radius distributions after applying synthetic observational filters. The sky-blue curve corresponds to the CKS sample.
\label{fig:RV_obs}}
\end{figure*}

Based on these synthetic transit projections, we compute the observed distributions of period ratio, radius ratio, and system multiplicity for super-Earths and mini-Neptunes with orbital periods $P<100 \ \rm d$ (Fig. \ref{fig:obs}), as well as the radius distribution around the observed ``radius valley'' (Fig. \ref{fig:RV_obs}). For context, we compare our results to the California-Kepler Survey (CKS) sample, shown as solid gray lines in Fig. \ref{fig:obs} and as sky-blue lines in Fig. \ref{fig:RV_obs}). Although reproducing the observed distributions in detail is not the goal of this study, we are interested in evaluating whether different disk evolution scenarios produce observable trends broadly consistent with the data. For the non-rebound scenario, we include only dynamically unstable systems, as they more closely resemble observed planetary system architectures.

The results show that the various disk evolution models yield broadly similar observable distributions, with only minor discrepancies relative to one another and to the CKS sample, consistent with the findings of \citet{Shibata&Izidoro2025}. This suggests that the magnetospheric rebound effect, while important for internal dynamics, has a limited impact on the statistical properties observable in transit surveys. We therefore focus our discussion on the most significant differences among scenarios. % between the intrinsic and observational distributions 

% Fig. \ref{fig:obs} illustrates the period ratio (panel a), radius ratio (panel b), and multiplicity (panel c) of transit-simulated planetary systems, where colored curves represent different formation models. 

% Compared with CKS samples, most compact first-order resonances (e.g. 6:5, 5:4, and 4:3) are disrupted in our simulations. 

As shown in Fig.~\ref{fig:models}a, the intrinsic period ratio distribution in the fiducial scenario is more dispersed than in the short-lived disk case. However, because short-lived disks tend to produce higher eccentricities and inclinations, their synthetic transit observations (Fig.~\ref{fig:obs}a) wash out the compactness of the system. As a result, the observed period ratio distributions for both scenarios appear nearly indistinguishable. The contrast between the long-lived disk model and other models also becomes less pronounced after taking observational biases into account, though it does not entirely disappear. 
Compared with the results of \citet{Shibata&Izidoro2025} (see the middle-left panel of their Fig. 13), the rebound scenario erodes most compact and fragile first-order resonant structures, such as the 6:5 and 5:4 commensurabilities.

In terms of observed multiplicity (Fig.~\ref{fig:obs}c), about 60-70\% of synthetic systems appear as single-transiting planets, 20–30\% as two-planet systems. And only a small fraction of systems contain three or more planets, in agreement with \citet{Shibata&Izidoro2025}. When combined with the intrinsic multiplicity results (Fig.~\ref{fig:multi}), this highlights the strong impact of geometric and detection biases: even systems with several planets often appear as singles due to limited transit visibility \citep{Izidoro2017}. All models show good agreement with the observed system multiplicity distribution, with only a slight excess in the fraction of multi-planet systems.

The bimodal distribution of planetary radii remains evident in the rebound scenarios (Fig.~\ref{fig:RV_obs}), with the relative prominence of the second peak growing as the disk recession timescale increases.
However, the second peak in the unstable case is notably diminished. This may contribute to the excitation of planetary eccentricities and inclinations after disk dispersal, which reduces their detection probability.
All scenarios, aligned with \citet{Shibata&Izidoro2025}, produce fewer small rocky planets ($R_{\rm p} < 1 \ R_{\oplus}$) and Neptune-like planets ($R_{\rm p} \sim 4 \ R_{\oplus}$) relative to observations.

% Since the transit probability scales with $R_{\star}/a_{\rm p}$ \citep{Borucki&Summers1984}, long-period icy planets are more difficult to detect. Thus, the second peaks in the observed radius distributions in Fig. \ref{fig:RV_obs} become less pronounced than in Fig. \ref{fig:RV}. Despite the nearly vanished second peak in the non-rebound scenario, our main conclusions remain robust.

% Overall, the observed period ratio distributions are highly consistent across different scenarios.

\section{Conclusion} \label{sec:summary}

In this paper, we conducted N-body simulations to explore the effect of the magnetospheric rebound mechanism on breaking the resonance chain between super-Earths and mini-Neptunes whose orbital period is shorter than 100 days. We investigated the dynamical evolution of planetary systems under various disk models, including different depletion timescales and the inclusion or exclusion of the rebound effect. Additionally, we analyzed both statistical and observational distributions of planetary orbits and physical characteristics. The main conclusions are summarized below.

\begin{itemize}
    \item The magnetospheric rebound effect plays a significant role in destabilizing planetary systems and disrupting their tightly packed first-order resonances. Super-Earths and mini-Neptunes are driven outward during cavity expansion, which significantly increases their mutual separations. As the system evolves, over 90\% of first-order resonances (e.g., 6:5, 5:4, 4:3, 3:2) are disrupted, leading to a significant reconfiguration of the planetary architecture. 

    \item Longer disk depletion timescales strengthen the rebound effect, leading to slightly larger final mutual separations of planets. As the disk dissipation timescale increases up to $5 \times 10^5$ yr, enhanced inward migration of outer icy planets promotes stronger orbital instabilities and more frequent mass doubling.
    
    \item Although slight differences remain, the intrinsic statistical distributions ---including period ratios, radius ratios, radius valley, and multiplicity---of unstable systems formed in the non-rebound scenario are comparable to those formed in rebound cases with disk dissipation timescale $t_{\rm disk} \le 10^5 \ \rm yr $. The discrepancies could be further diminished in synthetic transit observations.

Overall, magnetospheric rebound could be an effective mechanism for breaking resonant chains by triggering dynamical instabilities within super-Earth and mini-Neptune systems during cavity expansion. These instabilities, unlike those driven by gravitational or secular interactions, emerge early in the gas-rich phase of disk evolution. Future time-resolved observations of resonant structures are essential to constrain the dominant mechanisms. 

\end{itemize}

% Overall, the magnetospheric rebound effect is an efficient mechanism for breaking the stability of planetary systems and disrupting resonance chains.
\appendix
\counterwithin{figure}{section}

\section{Final structure of planetary systems} \label{app:sys}

\begin{figure}
\includegraphics[width=\columnwidth]{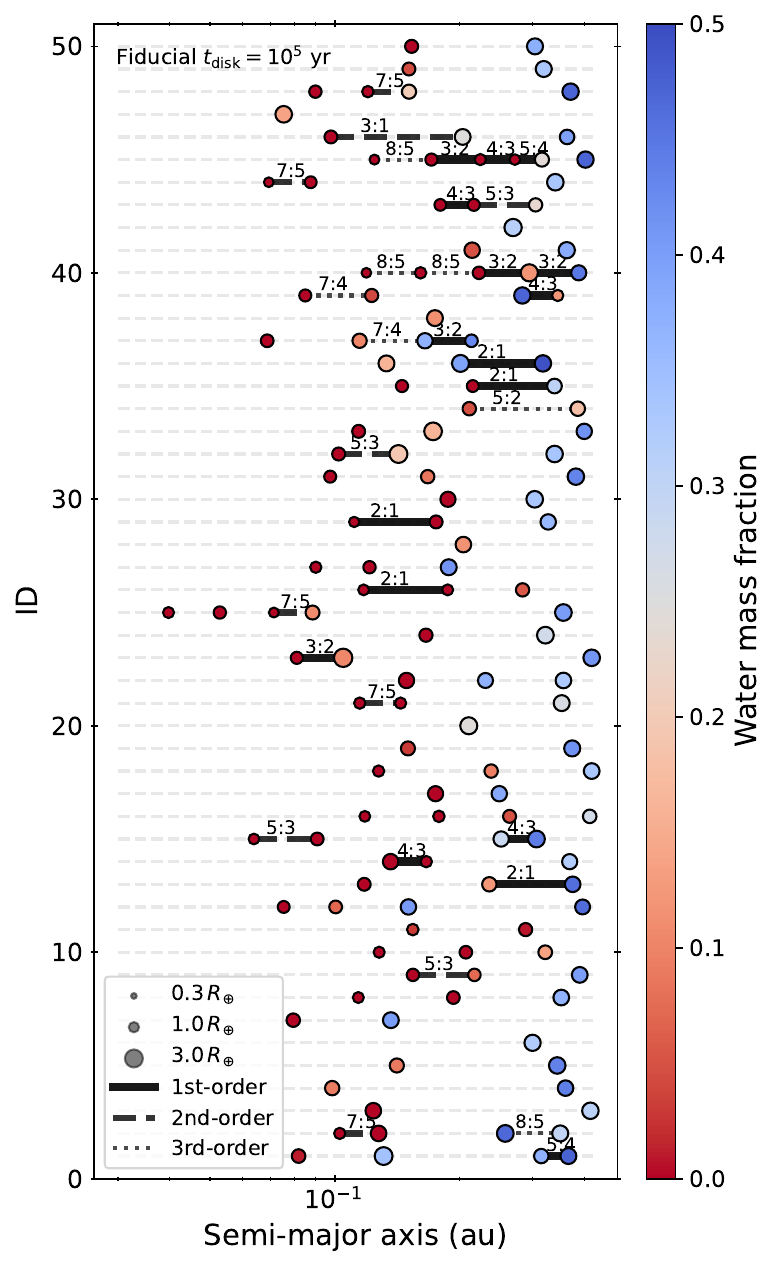}
\caption{Architecture of final planetary systems in the fiducial scenario. The size and color of dots correspond to the radius and water mass fraction of super-Earths/mini-Neptunes, respectively. The solid, dashed, and dotted black lines connecting adjacent planets indicate the near first-, second-, and third-order mean motion resonances, respectively, whose values are also illustrated above the lines. Only planets with orbital period $P < 100 \ d$ are shown here.
\label{fig:fid_sys}}
\end{figure}

\begin{figure}
\includegraphics[width=\columnwidth]{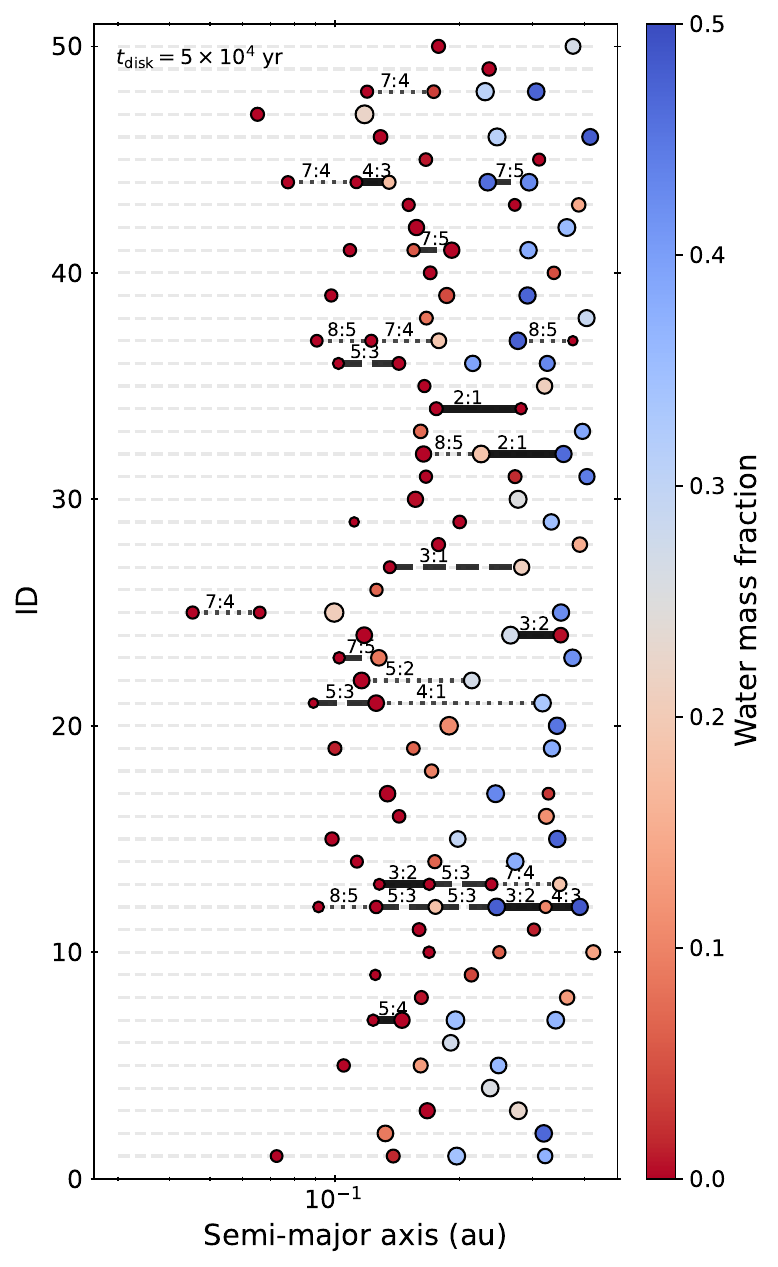}
\caption{Same as Fig. \ref{fig:fid_sys}, but for short disk lifetime cases.
\label{fig:short_sys}}
\end{figure}

\begin{figure}
\includegraphics[width=\columnwidth]{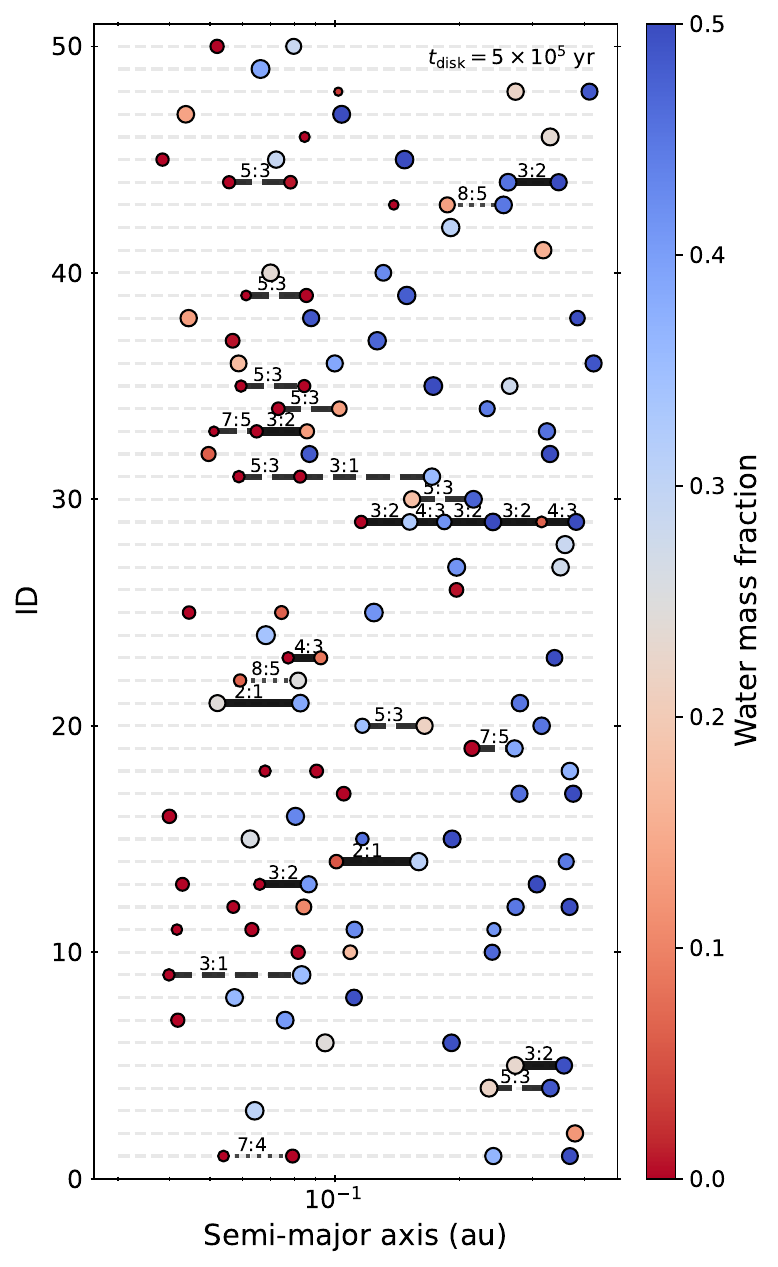}
\caption{Same as Fig. \ref{fig:fid_sys}, but for long disk lifetime model.
\label{fig:long_sys}}
\end{figure}

\begin{figure}
\includegraphics[width=\columnwidth]{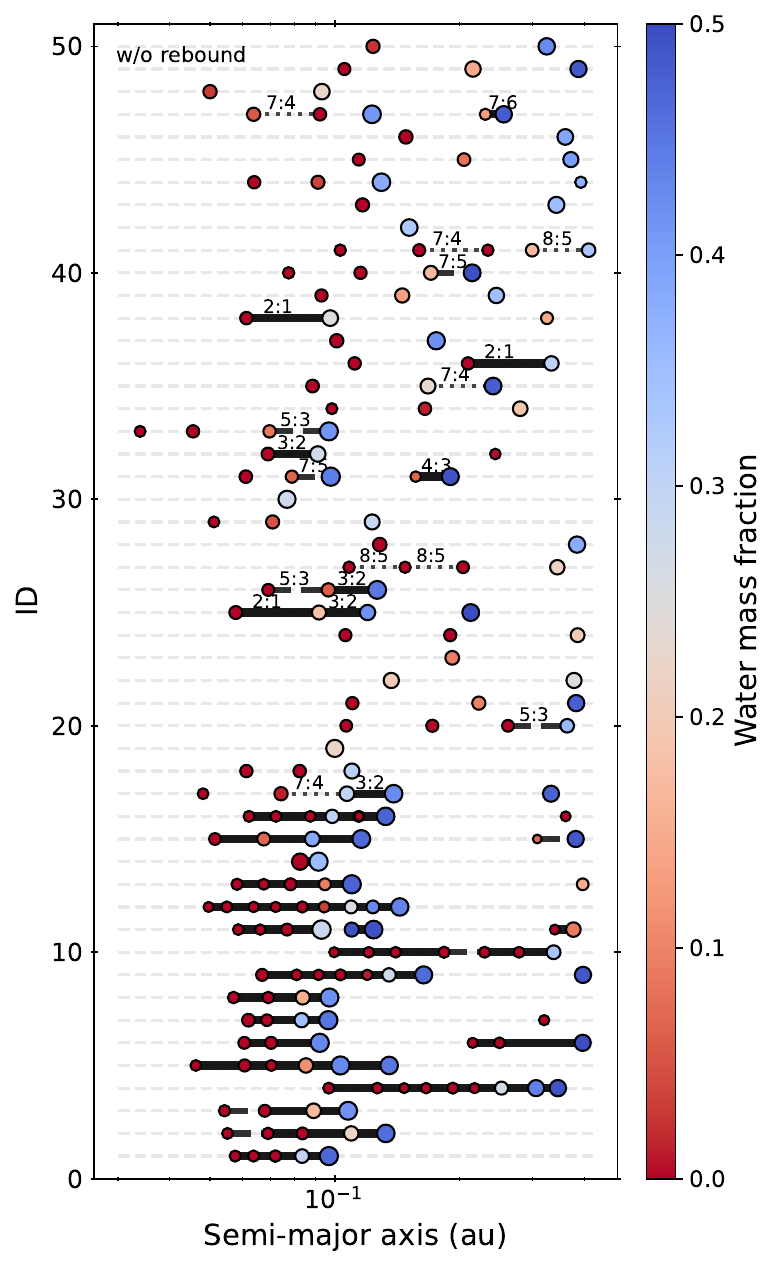}
\caption{Same as Fig. \ref{fig:fid_sys}, but for disks without rebound effect.
\label{fig:noreb_sys}}
\end{figure}

In this section, we examine the final orbital architectures of planetary systems formed under different disk evolution scenarios, as shown in Figs.~\ref{fig:fid_sys}--\ref{fig:noreb_sys}. Planets are represented as filled circles, color-coded by water mass fraction and scaled in size according to planetary radius. Pairs of planets located near mean-motion resonances---defined as having a fractional deviation $\delta < 0.03$ from exact commensurability---are connected by lines, with solid lines indicating first-order resonances and dashed lines representing second-order or higher order resonances. 

In the fiducial model (Fig.~\ref{fig:fid_sys}), the number of resonant pairs declines with increasing resonance order. Two systems (ID02 and ID15) exhibit multiple isolated resonance pairs, while four systems (ID37, ID40, ID43, and ID45) host resonant chains involving three or more planets. The longest chains, found in ID40 and ID45, link five planets in succession. In an evolution reminiscent of Kepler-80, the innermost planet in ID37 detaches from the chain following a dynamical instability.

In the short-lived disk scenario (Fig.~\ref{fig:short_sys}), a Trappist-1-like configuration emerges in system ID12, with compact, low-mass planets captured into successive second- and first-order resonances. The long-lifetime disk model (Fig.~\ref{fig:long_sys}) also supports extended resonant chains, with system ID29 forming a tightly packed 18:12:9:6:4:3 sequence, enabled by the rapid inward migration of outer icy planets.

In the non-rebound scenario (Fig.~\ref{fig:noreb_sys}), approximately 40\% of systems retain long resonant chains, consistent with the fraction that remain dynamically stable. These chains are typically composed of first-order commensurabilities, with the longest sequence linking nine super-Earths and/or mini-Neptunes.

In summary, systems formed without rebound tend to preserve more and longer first-order resonant chains. By contrast, in rebound scenarios, the outward expansion of disk's inner edge frequently disrupts these primordial configurations. This disruption reduces the prevalence of first-order resonances and increases the likelihood of forming second-order resonances as planets dynamically reorganize during and after instability. 

\newpage

\begin{acknowledgments}
We appreciate the constructive feedback from the anonymous referee. M.P., A.I., and S.S. acknowledge the Center for Origins and Habitability of the Rice Space Institute. This work was supported in part by the Big-Data Private-Cloud Research Cyberinfrastructure MRI-award funded by NSF under grant CNS-1338099 and by Rice University’s Center for Research Computing (CRC).
\end{acknowledgments}

\end{document}